\documentclass[%
 reprint,
nofootinbib,
 amsmath,amssymb,
 aps,
 prl,
]{revtex4-1}

\usepackage{float}
\usepackage[colorlinks = true, citecolor = blue, linkcolor = blue ]{hyperref}
\usepackage{graphicx}
\usepackage{dcolumn}
\usepackage{bm}

\newcommand{\jcap}{JCAP}

\newcommand{\apjl}{ApJL}

\newcommand{\aap}{A\&A}

\def \jcap {JCAP}
\def \aap {A\&A}

\def \qp {q_{\rm p}}
\def \qe {q_{\rm e}}
\def \tesc {\tau_{\rm esc}}
\def \tloss {\tau_{\rm loss}}
\def \ppbar {\bar{\text{p}}/\text{p}}
\def \epbar {\bar{p}/e^+}
\def \epspbar {\epsilon_{\bar{\text{p}}}}
\def \utot {u_{\rm tot}}
\def \ng {n_{\rm g}}
\def \etaB {\eta_{\rm B}}
\def \nB {n_{\rm B}}
\def \nC {n_{\rm C}}
\def \np {n_{\rm p}}
\def \npbar {n_{\bar{\text{p}}}}

\begin{document}

\preprint{APS/123-QED}

\title{Nonsecondary origin of cosmic ray positrons}

\author{Rebecca Diesing}
\author{Damiano Caprioli}%
\affiliation{%
Department of Astronomy and Astrophysics, University of Chicago, Chicago, IL 60637, USA}%

\begin{abstract}
We investigate the possibility of secondary production as the source of cosmic ray (CR) positrons, taking into account observed electron and antiproton spectra and allowing that the reported steepening in the AMS-02 positron spectrum may be due to radiative losses. Using a simple argument based on the antiproton spectrum, we show that positrons cannot be purely secondary. Namely, the antiproton to positron ratio requires that positrons be uncooled up to a few hundred GeV. This requirement implies that CR protons escape from the Galaxy too quickly to reproduce the observed antiproton flux. We further show that this result rules out also more complex secondary production scenarios such as reacceleration and the Nested Leaky Box. Thus, we conclude that while antiprotons can be produced exclusively in secondary interactions, a primary source of CR positrons ---be it dark matter or pulsars--- is required to explain their spectrum.

\end{abstract}

\maketitle 

\section{Introduction}
Understanding the origin of Galactic cosmic rays (CRs) requires a complete model of their acceleration and transport in the Galaxy. 
In the standard picture, nuclei and electrons are accelerated at the forward shocks of supernova remnants (SNRs), which provide sufficient energetics and an efficient acceleration mechanism, \emph{diffusive shock acceleration} (DSA) \citep[]{hillas05, caprioli+10a}. 
DSA predicts power-law energy distributions of CRs, $\propto E^{-q}$, with $q$ set by shock hydrodynamics \citep[]{krymskii77, axford+77p, bell78a, blandford+78}. 

CRs subsequently propagate through the Galaxy with an energy-dependent residence time (depending, e.g., on the scaling of the diffusion coefficient with energy), resulting in modification of their spectra. 
Electrons also lose energy via synchrotron and inverse-Compton scattering, resulting in further spectral modification. 
Thus, if SNRs inject protons with a spectrum that goes as $E^{-\qp}$ and their residence time, $\tesc$, has an energy dependence $\tesc \propto E^{-\delta}$, we expect the observed proton spectrum to go as $E^{-(\qp+\delta)}$. 
Similarly, if electrons are injected with $N_{\rm e} \propto E^{-\qe}$, we expect an observed electron spectrum $\propto E^{-(\qe + \beta)}$, where $\beta\gtrsim\delta$ encompasses spectral steepening due to energy loss and escape. At low energies, where $\tesc$ is much shorter than the loss time, $\tloss$, $\beta \simeq \delta$. At high energies, where $\tloss \ll \tesc$, $\beta = (1+\delta)/2$ \citep[e.g.,][]{amato+17, lipari19a}.

CR hadrons also interact with nuclei in the interstellar medium (ISM) to produce secondary particles, notably positrons, antiprotons, and spallation products such as boron and beryllium. 
In this picture, secondaries will have spectra $\propto E^{-(\qp +2\delta)}$. Thus, hadronic secondary-to-primary ratios such as boron over carbon (B/C) and the antiproton to proton ratio ($\ppbar$) go as $E^{-\delta}$. 

This picture breaks down in the spectrum of positrons. By the above logic, the positron spectrum goes as $E^{-(\qp + \delta + \beta)}$, meaning that the positron fraction, $e^+/(e^+ + e^-)$, should go as $E^{\qp - \qe + \delta}$, since the positron flux is much smaller than the electron flux such that $e^+/(e^+ + e^-) \simeq e^+/e^-$. 
In the approximation that $\qp = \qe$, the positron fraction thus decreases as $E^{-\delta}$. Yet, PAMELA and AMS-02 report a positron fraction that rises roughly as $E^{0.3}$ \citep[]{pamela13, ams14}. 
This ``positron excess" has been cited as evidence for a primary source of positrons, notably pulsars \citep[e.g.,][]{busching+08, hooper+09} and dark matter annihilation \citep[e.g.,][]{cholis+09} \cite[see][for a thorough discussion of potential positron sources]{serpico12}. 
More recently, AMS-02 reported a steepening in the positron spectrum around $\sim 300$ GeV, which they attribute to a primary source cutoff \citep[]{ams19a}.

Intriguingly, however, the positron to antiproton ratio is consistent with proton-proton branching ratios \citep[e.g.][]{lipari17,lipari19b,blum+18}, as expected if positrons are secondaries. More substantively, mitigating factors raised in the literature suggest that the observed positron flux may yet be explained by secondary production. First, a portion of the rise in the positron fraction likely arises from the fact that $\qp < \qe$; electrons experience synchrotron losses in the amplified magnetic fields of SNRs such that $0.1 < \qe - \qp < 0.4$ \citep{diesing+19}. Second, secondary positrons are likely injected with a harder spectrum than their parent protons due to the fact that a) an increase in multiplicity leads to a rise with energy in the effective positron production cross section \citep[]{korsmeier+18}, b) positrons are produced with $\simeq10\%$ the energy of their parent protons, meaning that their spectrum reflects the proton spectrum at 10$\times$ higher energies, where it is harder \citep[]{CALET19, ams15a, atic09, cream10_coll}, and c) some positrons are produced by heavier CR nuclei, which have harder spectra than hydrogen \citep[]{ams15b, cream11, pamela11}. The combined effect of the steepened electron spectrum and the hardened positron spectrum may be sufficient to reproduce a positron fraction with a slope of 0.3. As for the steepening in the positron spectrum at high energies, synchrotron and inverse Compton losses may account for such a rollover.

Furthermore, a host of more complex propagation models have been presented as resolutions to the positron excess that do not require a primary source \citep[e.g., ][]{cowsik+14, dogiel+90, malkov+16}. In general, these solutions rely upon inhomogeneity in CR diffusion which can introduce additional degrees of freedom to the problem.

In this paper, we will investigate the secondary origin of CR positrons in light of the new AMS-02 positron data \citep[]{ams19a}. We will show that it is possible to simultaneously fit the slopes of the observed positron, electron and antiproton spectra assuming only secondary production of positrons and antiprotons. However, using a novel estimate of the CR residence time based on the featureless antiproton to positron ratio and the antiproton flux normalization, we will show that the AMS-02 data rules out the purely secondary origin of positrons. We will also--for the first time--use this argument to rule out the more exotic secondary origin scenarios mentioned above.

Our argument proceeds as follows: to provide good agreement with observations, electrons and positrons must be uncooled below $\simeq 250$ GeV. 
For conservative estimates of Galactic magnetic fields and photon backgrounds, this requirement demands a short escape time from the Galaxy: $\tesc \simeq 0.66$ Myr at 250 GeV. 
This $\tesc$ is incompatible with B/C and $\ppbar$, as such fast escape of primaries leaves insufficient time to produce the observed secondary flux given acceptable ISM densities.

\section{Method}
To test the validity of the secondary production scenario, we perform a simple calculation of the antiproton, electron, and positron spectra using the CALET proton spectrum \citep[]{CALET19}, assuming no primary contribution of antiprotons or positrons. The CALET data are chosen for their large range of energies (50 GeV -- 10 TeV). 

The following outlines our calculation technique for each species. In all cases, spectra are arbitrarily normalized to match observations; our concern here is the spectral slope. 
We consider only CRs above $E \simeq 20$ GeV, where solar modulation is negligible. 
More generally, our approach is designed to be as generous as possible towards the secondary positron production scenario in order to convincingly rule it out later in this paper.

Antiprotons: In the secondary production scenario, antiprotons are produced in interactions between CR and ISM nuclei with $\sim10\%$ the energy per nucleon of the parent CR \citep[]{korsmeier+18}. Thus, at production, the antiproton spectrum follows that of protons with energies 10 times larger. This prediction gives an antiproton injection spectrum that is harder than that of the parent protons, since the proton spectrum hardens at high energies \citep[]{CALET19, ams15a, atic09, cream10_coll}. Furthermore, due to the fact that a) the antiproton production cross section rises with energy \citep{korsmeier+18} and b) a fraction of antiprotons are produced by heavier CR nuclei which have harder spectra than protons \citep[]{ams15b, cream11, pamela11}, we can expect an additional hardening of the antiproton spectrum, herein parametrized by $E^{\epspbar}$.

To estimate the antiproton spectrum, we begin by mapping the CALET proton spectrum onto energies that are ten times smaller. We then harden this spectrum according to the results of \cite{korsmeier+18}, which calculates the antiproton source term accounting for cross sectional effects and the contribution of heavier elements. \cite{korsmeier+18} predicts a spectral hardening that asymptotically approaches $\epsilon_{\bar{p}} \simeq 0.15$.

Finally, to account for propagation, the antiproton spectrum is steepened by $\delta$. The result is compared to observations, and $\delta$ is chosen to produce good agreement with the antiproton spectrum and B/C.

Electrons: Like protons, electrons are accelerated at the forward shocks of SNRs via DSA. Since DSA is rigidity dependent, the electron spectrum at injection is expected to follow that of protons, modulo steepening due to synchrotron losses in the amplified magnetic fields of SNRs. This steepening, parametrized by $\Delta q = \qe - \qp$ is estimated to vary between 0.1 and 0.4, depending on the CR acceleration efficiency and the density of the circumstellar medium \citep[]{diesing+19}.

At low energies, CR electrons experience negligible radiative (synchrotron and inverse Compton) losses during propagation through the Galaxy, since $\tloss \gg \tesc$. At such energies, spectral steepening during propagation is solely due to diffusion and we therefore estimate the electron spectrum at low energies as the CALET proton spectrum steepened by $\Delta q$. 

At high energies, $\tau_{\rm loss} < \tau_{\rm esc}$ and energy losses are non-negligible. The result is a spectral steepening of $(1+\delta)/2$ relative to the spectrum at injection \citep[e.g.,][]{amato+17, lipari19a}. Since the observed proton spectrum has already been steepened by $\delta$ relative to the injection spectrum, we calculate the observed electron spectrum at high energies as the CALET proton spectrum steepened by $\Delta q + (1-\delta)/2$.

The transition between the low and high-energy regimes occurs at $E^*$, where $\tau_{\rm loss} = \tau_{\rm esc}$. Estimating $\tloss$ is relatively straightforward, as it depends only on $E$ and $\utot$, the total energy density in magnetic and radiation fields. From \cite{lipari19b} we have

\begin{equation}\label{eq:tloss}
    \tloss \simeq 310.8\bigg(\frac{\text{eV cm}^{-3}}{\utot}\bigg)\bigg(\frac{\text{GeV}}{E}\bigg) \text{Myr}.
\end{equation}

In this analysis we assume a typical galactic magnetic field of a few $\mu$G ($u_{\rm B} \simeq 0.5 \text{ eV cm}^{-3}$ and a total radiation energy density $u_{\rm rad} \simeq 1.4 \text{ eV cm}^{-3}$ \citep[see][for a detailed discussion]{draine11}, which yield $\tloss \simeq 164 \text{ Myr }E_{\rm GeV}^{-1}$. Note that the vast majority of the light contributing to $u_{\rm rad}$ comes from the cosmic microwave background, dust emission, and IR starlight with frequencies low enough that the Klein-Nishina modification to the Thomson cross section remains small at high CR energies. 

Meanwhile, $\tesc \propto E^{-\delta}$ with a normalization that can be constrained observationally. Arguably the best constraint on this normalization comes from radioactive clocks, namely $^{10}$Be. 
While direct measurements of $^{10}$Be are available only at low energies \citep[]{garcia-munoz+77, ahlen+00, yanasak+01}, the Be/B ratio can be used to estimate the $^{10}$Be flux, suggesting that $\tesc \sim 100$ Myr at 10 GeV \citep[]{evoli+19b}. 
Still, some models of CR transport argue that antimatter might experience a shorter escape time than spallation products like beryllium. For example, the Nested Leaky Box posits that, at a few tens of GeV, spallation products are largely produced and confined in cocoons near CR sources, while antimatter is produced throughout the Galactic volume, where $\tesc$ may be small \citep[see, e.g.,][as well as our discussion later in this paper]{cowsik+14}. For this reason, we leave $\tesc$ and thus $E^*$ as a free parameter and use CR lepton spectra to place constraints. 
We will discuss the implications of these constraints later in the paper, as they are crucial to ruling out the secondary origin of CR positrons. 

Positrons: Like antiprotons, positrons are produced in interactions between CR and ISM nuclei with approximately 10\% the energy per nucleon of the parent CR. They, too, are expected to experience some spectral hardening, $\epsilon_{e^+}$, due to multiplicity effects and the contribution of helium. 
Since $\epsilon_{e^+}$ has not been calculated explicitly, we take it to be equal to $\epsilon_{\bar{p}}\simeq 0.15$ as calculated in \cite{korsmeier+18}. 
Note that this is the most favorable limit for the secondary positron origin, since, in general, $\epsilon_{e^+}\lesssim \epsilon_{\bar{p}}$.

Thus, at low energies ($E < E^*$), the positron spectrum is estimated as the proton spectrum at energies 10 times higher, hardened by $\epsilon_{\bar{p}}$ and steepened by $\delta$ to account for propagation. At high energies ($E > E^*$), radiative losses become non-negligible and the positrons are instead steepened by $(1+\delta)/2$. This steepening may account for the apparent high-energy cutoff in the positron spectrum reported by AMS-02 \citep{ams19a}.

\section{Results}
\begin{figure}[h]
    \centering
    \includegraphics[trim=5px 20px 30px 50px, clip, width=0.49\textwidth]{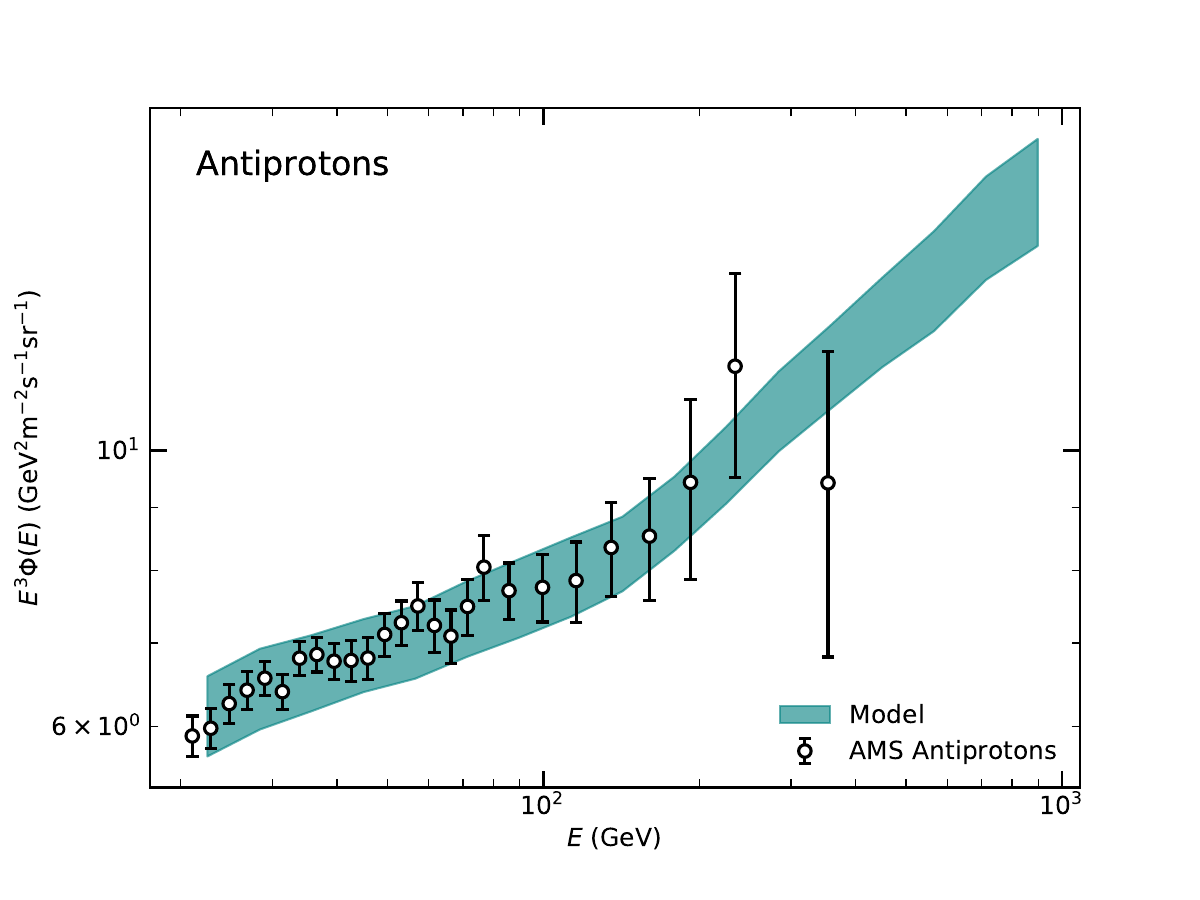}
    \caption{Predicted CR antiproton spectrum (teal band) overlaid with AMS-02 data \citep[]{ams16a}, assuming only secondary production of antiprotons. The width of the band is determined by propagating uncertainties on the CALET proton data \citep{CALET19} used to generate this prediction. We take $\delta = 0.4$ for $E<200$ GeV and $\delta = 0.2$ for $E>200$ GeV, which yield good agreement between predicted and observed spectra.}
    \label{fig:antiprotons}
\end{figure}

Figure \ref{fig:antiprotons} shows predicted and observed antiproton spectra, calculated as described above in order to constrain $\delta$. $\delta = 0.4$ for $E<200$ GeV and $\delta = 0.2$ for $E>200$ GeV yield a good fit, and are consistent with B/C \citep[]{ams18}. Physically speaking, a broken $\delta$ may be the result of CR-induced turbulence in the Galaxy \citep[e.g.,][]{blasi+12}. While the large error bars on the antiproton spectrum provide relative freedom in our choice of $\delta$, any $\delta\gtrsim 0.5$ can be safely ruled out.

\begin{figure}[h]
    \centering
    \includegraphics[trim=5px 50px 40px 70px, clip, width=0.49\textwidth]{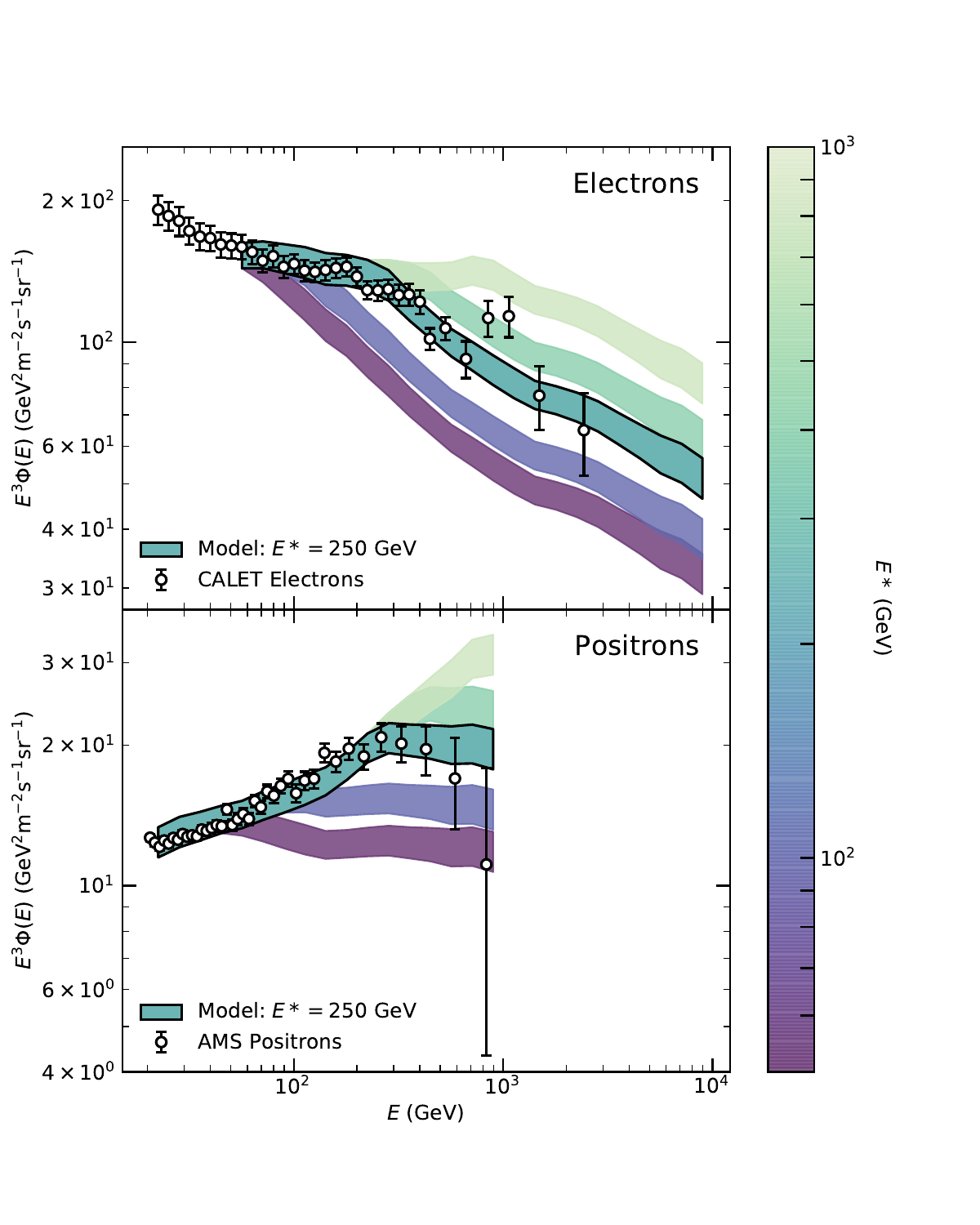}
    \caption{Predicted CR lepton spectra overlaid with CALET (electron \citep[]{calet17}) and AMS-02 (positron \citep[]{ams19a}) data assuming only secondary production of positrons. The color scale denotes the value taken for $E^*$, the energy where escape and loss times are equal. In all models, we take $\delta$ as in Figure \ref{fig:antiprotons} and $\Delta q = 0.25$. $E^*\simeq250$ GeV yields a reasonably good fit to both electron and positron data.}
    \label{fig:leptons}
\end{figure}

Figure \ref{fig:leptons} shows predicted CR electron and positron spectra for various values of $E^*$, assuming no primary origin of positrons and taking $\delta$ as above. 
For $E^* \simeq 250$ GeV, the predicted spectra are in fairly good agreement with both CALET electron \citep[]{calet17} and AMS-02 positron data \citep[]{ams19a}, while lower values of $E^*$ fit poorly. Admittedly, the steepening in the observed positron spectrum appears to be slightly larger than the $(1-\delta)/2$ predicted for the transition to a loss-dominated energy regime. However, given the large error bars on the AMS-02 data, this discrepancy may not be real. It is also worth noting that some experiments suggest a featureless electron spectrum up to a break around 1 TeV \citep[e.g.,][]{dampe17}. This would imply either a much higher value of $E^*$ (unlikely, given the results from this work), or a break in the electron source term at TeV energies (which is quite reasonable \citep[]{diesing+19}). Regardless, in the secondary production scenario, $20$ GeV $\lesssim E^* \lesssim 250$ GeV can be safely ruled out by the existing lepton data.

The main takeaway from this result is that, in order to produce good agreement with data, positrons must be uncooled up to at least 250 GeV. While the relatively featureless electron spectrum can also accommodate much smaller values of $E^*$ ($\lesssim 3$ GeV, as discussed in \cite{lipari19a}) or could be impacted by an additional nearby source of primary electrons \cite[e.g.,][]{manconi+19}, the positrons, if purely secondary, have no such flexibility. This can be seen most explicitly in the antiproton to positron ratio, $\epbar$. \cite{ams16a} shows that $\epbar$ is roughly constant in energy from a few tens of GeV up to a few hundred GeV. Since there is no physical reason for secondary positrons to be produced with a harder spectrum than antiprotons, an energy-independent $\epbar$ implies that, if positrons are purely secondary, they cannot be cooled.

\section{Discussion}
Let us now consider the implications of uncooled leptons, namely, of $E^* \gtrsim 250$ GeV. 
For $\tloss \simeq 164\text{ Myr }E^{-1}_{\rm GeV} $ (Equation \ref{eq:tloss}), $\tesc = \tloss \simeq 0.66$ Myr at 250 GeV. 
Here we make no assumptions about the value of $\delta$. Let us also consider CR boron, which is produced via spallation of carbon along with other nuclei that have an atomic number $> 5$.
The boron production rate per unit volume is thus $\etaB(E) \simeq \ng\nC(E)\sigma_{\rm B}(E) c$, where $\ng$ is the average number density of nucleons in the galactic halo, $\nC(E)$ is the number density of CR carbon, and $\sigma_{\rm B}(E)$ is the total boron production cross section at energy per nucleon $E$. B/C is thus given by

\begin{equation}
    \frac{\nB(E)}{\nC(E)} \simeq \frac{\etaB(E) \tesc(E)}{\nC(E)} = \ng \sigma_{\rm B}(E) c \tesc(E).
\end{equation}

Since B/C is a known quantity, we can estimate the average number density required to reproduce observations,

\begin{equation}
    \ng \simeq \frac{\nB(E)}{\nC(E)}\frac{1}{\sigma_{\rm B}(E)c\tesc(E)}.
\end{equation}

At $E = 250$ GeV, B/C $\simeq 0.1$ \citep{ams18}. If we take $\sigma_{\rm B}(250 \text{ GeV}) \simeq 120$ mb \citep[]{webber+03, evoli+19a}, $\tesc(250 \text{ GeV}) \simeq 0.66$ Myr gives $\ng \simeq 1.3 \rm{\ cm^{-3}}$. Note that this is the average nucleon density over the entire galactic halo required to produce the observe boron flux. If we assume that the Galaxy is composed of a thin disk with density $n_{\rm disk}$ and scale height $h$, and a diffuse halo with density near zero and scale height $H$, we find that $n_{\rm disk} = \ng H/h$. For canonical values of $h \sim 300$ pc and $H \sim 6$ kpc, numbers that are consistent with the results of the B/C analysis in \cite{evoli+19b}, $n_{\rm disk} \sim 26 \text{ cm}^{-3}$. Obviously, this density is far too high to be acceptable, given that the average hydrogen number density in the disk is only $\sim 1 \text{ cm}^{-3}$ \citep[]{ferriere+01}. 
Moreover, since this analysis does not consider boron destruction via spallation, it is a significant underestimate of the required $\ng$. In short, if carbon and positrons have the same residence time in the Galaxy, positrons cannot be secondary. 

However, it has been suggested that $\tesc$ may be longer for spallation products than for secondary antimatter. 
For example, in the Nested Leaky Box (NLB) scenario \citep[]{cowsik+14}, primary CRs are accelerated in sources surrounded by cocoon-like regions where the confinement time is strongly energy-dependent (i.e., $\delta$ is large). Outside of these regions, the confinement time is small and $\delta \simeq 0$. 
Thus, primary CRs with energies below a few hundred GeV would remain confined within cocoons for a long time, giving them a large overall $\tesc$ (time in the cocoon + time in the halo). Above a few hundred GeV, primary CRs would quickly escape from cocoons and spend most of their overall confinement time in the halo. 
Since secondary hadrons are produced with the same energy per nucleon as their parent CRs, those with energies under few hundred GeV per nucleon would be primarily produced inside cocoons. 
Positrons and antiprotons, on the other hand, are produced by protons with roughly ten times their energy, meaning that these parent protons quickly escape from cocoons, and antimatter with $E$ greater than a few tens of GeV must be produced in the ISM. Thus, in the NLB model, secondary antimatter probes a shorter escape time than spallation products. 

Furthermore, one could argue that carbon experiences acceleration and transport that somehow differs from that of protons, perhaps because carbon might be accelerated at the SNR reverse shock (which propagates in the carbon-rich ejecta), or because different SNR classes (namely, core-collapse and Type Ia) might preferentially accelerate and/or confine carbon. 
Thus, to truly rule out secondary production as the primary source of positrons, we must consider not only boron production, but also antiproton production. Since antiprotons and positrons are produced primarily in proton-proton interactions with 10\% the energy of their parent nucleon, their escape times cannot be decoupled.

Repeating the analysis we performed with B/C, we have

\begin{equation}
   \ng \simeq \frac{\npbar(E)}{\np(E)}\frac{\np(E)/\np(10E)}{\sigma_{\rm \bar{p}}(E)c\tesc(10E)}.
\end{equation}

Note that, for antiprotons, $\ng$ depends on the CR proton number density and $\tesc$ at energies that are ten times $E$, the energy of the antiproton (and the energy at which we measure $\ppbar$). 
From \cite{ams16a}, $\ppbar \simeq 2.2\times10^{-4}$ at $E \simeq 250$ GeV. $\tesc(2.5 \text{ TeV})$ required for $E^* \simeq 250$ GeV is simply $0.66 10^{-\delta}$ Myr, with $\delta = 0.2$ as discussed previously. 
Meanwhile, \cite{korsmeier+18} gives $\sigma_{\rm \bar{p}}(250 \text{ GeV}) \lesssim 19$ mb for proton-proton interactions alone. Based on the full results of \cite{korsmeier+18}, we multiply this cross section by a factor of 1.8 to account for other production channels. Note that 19 mb is the 250 GeV antiproton production cross section for parent protons with very high energies, $ E \gg 2.5$ TeV. Since the antiproton production cross section increases with the energy of the parent particle, this choice of cross section ensures that we do not underestimate antiproton production.
Taking the observed proton spectrum as $\propto E^{-2.7}$ so that $\np(E)/\np(10E) \simeq 500$, we find $\ng \simeq 8.3 \text{ cm}^{-3}$, far too large to be feasible. 

Thus, we rule out NLB and, more generally, any secondary positron production model that decouples $\tesc$ for spallation products from $\tesc$ for antimatter. Note that \cite{cowsik+16} reached a different conclusion because they assumed $\tesc \gtrsim 2$ Myr, corresponding to $E^* \simeq 80$ GeV, far too low to reproduce the latest measurement of $\epbar$.

This argument also rules out reacceleration at the source as an explanation for the positron excess \citep[e.g.,][]{blasi09,blasi+09}. In this picture, secondary positrons and antiprotons are produced inside CR sources (SNRs), with more energetic secondaries being preferentially reaccelerated because of their longer diffusion length. While this effect flattens both positron and antiproton spectra, thereby providing a rising positron fraction, it still requires that positrons be uncooled up to a few hundred GeV in order to be consistent with an energy-independent $\epbar$.

A final class of propagation models invokes molecular clouds as sources of additional grammage and thus secondary production \cite[e.g., ][]{dogiel+90, malkov+16}. These models suggest that, with fine-tuned molecular cloud density, size, and magnetic field, it may be possible to confine primaries such that they produce sufficient secondaries without cooling positrons. However, keeping the positrons uncooled requires an extremely short residence time, $10^4 - 10^5$ years, assuming cosmic rays spend all of this time in molecular clouds. This is inconsistent with $^{10}$Be/$^9$Be measurements, which suggest an escape time $\gtrsim 10$ Myr at a few hundred GeV \citep[]{evoli+19b}. If molecular clouds are the dominant source of grammage in the Galaxy, we would expect the escape time probed by radioactive clocks to be much lower. More generally, this argument implies that inhomogeneous diffusion does not resolve the positron excess.

\section{Conclusion}

At high energies, most CR positrons cannot be of secondary origin. Taken together, positron and antiproton spectra require that positrons be uncooled up to a few hundred GeV. Given the known radiation and magnetic fields in the Galaxy, this requires an escape time $\lesssim 0.66$ Myr at $E \simeq 250$ GeV. This escape time is incompatible with observations, as it requires an infeasibly large average gas density in the halo to produce the observed boron and antiproton flux. Furthermore, this argument rules out reacceleration, NLB, molecular cloud confinement, and other secondary production models that rely on inhomogeneous diffusion or decoupling spallation products from antimatter.

Thus, we must conclude that a primary source of positrons, be it dark matter or astrophysical, dominates their flux above $\sim 10$ GeV.

\begin{acknowledgments}

\section{acknowledgments}
We thank Michael Korsmeier, Fiorenza Donato, Pasquale Blasi, and Elena Amato for their comments and discussions on secondary particle production. We also thank Hsiao Wen Chen and Erin Boettcher for their discussion on ISM density.  This research was partially supported by a Parker Fellowship, NASA (grant NNX17AG30G, 80NSSC18K1218, and 80NSSC18K1726) and the NSF (grants AST-1714658 and AST-1909778).
\end{acknowledgments}

\end{document}